\documentclass[aip,pof]{revtex4-1}

\usepackage{graphicx}
\usepackage{dcolumn}
\usepackage{bm}
\usepackage{psfrag}

\begin{document}

\preprint{APS/123-QED}

\title{High Rayleigh number convection with double diffusive fingers}

\author{E. Hage}

\author{A. Tilgner}

\affiliation{Institute of Geophysics, University of G\"ottingen,
Friedrich-Hund-Platz 1, 37077 G\"ottingen, Germany }

\date{\today}

\begin{abstract}
An electrodeposition cell is used to sustain a destabilizing concentration difference of
copper ions in aqueous solution between the top and bottom boundaries of the
cell. The resulting convecting motion is analogous to Rayleigh-B\'enard
convection at high Prandtl numbers. In addition, a stabilizing temperature
gradient is imposed across the cell. Even for thermal buoyancy two orders of
magnitude smaller than chemical buoyancy, the presence of the weak stabilizing
gradient has a profound effect on the convection pattern. Double diffusive
fingers appear in all cases. The size of these fingers and the flow velocities
are independent of the height of the cell, but they depend on the ion
concentration difference between top and bottom boundaries as well as on the
imposed temperature gradient. The scaling of the mass transport is compatible
with previous results on double diffusive convection.
\end{abstract}

\maketitle

\section{Introduction}

Double diffusive convection occurs in a fluid whose density depends on two
properties with different diffusion constants. The prototypical example is water
whose density depends on temperature and salt concentration. This particular
combination is relevant in oceanography and the interest in double diffusive
convection historically originated in this field. More applications have since
emerged in geophysics, astrophysics, and engineering \cite{Turner85}. Double
diffusive convection differs from ordinary Rayleigh-B\'enard convection in that
two fundamentally new flow structures can appear: Layers and fingers. Fingers
are vertically oriented long and narrow regions of up- or downwelling fluid
which can occur if the fluid property with the large diffusion coefficient (for
example temperature) imposes a stabilizing gradient on the fluid, whereas the
property with the small diffusion constant (the salt concentration in the
oceanographic example) is unstably stratified. The basic principles of finger
formation are well understood since linear stability analysis already predicts
convection with fingers \cite{Baines69}. Layers dividing the fluid volume into several
sections in the vertical direction with little advective transport between the
layers is frequently observed in double diffusive systems, irrespective of which
component is stabilizing and which is destabilizing. The necessary conditions
for layer formation remain controversial \cite{Stern69, Merryf00, Radko2005}.

The experiments reported here deal with a double diffusive convective system
which forms fingers but no layers so that various scaling laws obeyed in the
finger regime can be determined accurately.

A popular experimental setup for the study of double diffusive convection has
been water with gradients of salt and sugar concentrations. The diffusion
coefficients of salt and sugar differ by a factor of about 3. However, the ratio
of the diffusivities of temperature and salt is much larger, typically around 100.
The large aspect ratio of fingers
makes their numerical simulation difficult. Experiments on the other hand have a
problem in maintaining a steady state. Many experiments start from an initial
distribution of salt and temperature and let the salt stratification disappear
in the course of time. Some results may then depend on whether an experiment was
started from a continuous stratification or a step function in the salt
distribution \cite{Sreeni09}. One is therefore interested in experiments capable
of keeping both constant temperature and concentration differences between top
and bottom boundaries. Maintaining a temperature difference is simple, but the
salt stratification is problematic. One solution is to use an apparatus with
permeable membranes as top and bottom boundaries \cite{Krishn03,Krishn09}. Water
tanks with constant temperature and salt concentration placed outside the
convecting volume on the other side of each membrane guarantee steady conditions
in the experimental cell.

An alternative solution to the problem consists in using an electrochemical
system. Convection in electrochemical cells with vertical electrodes has been
studied at least since 1949 \cite{Wagner49, Wilke53}. Even 
an application to double diffusion with horizontal gradients appeared \cite{Kamota85}.
Electrochemical convection was later noted in a cell with horizontal electrodes
\cite{Ward84} and developed into an
analogy for Rayleigh-B\'enard convection \cite{Goldst90}.
The principle of the system used most frequently
and also in this paper is to fill a cell with a solution of $CuSO_4$
and to place copper electrodes at the top and bottom of the cell. When an
electric current is sent through the cell, copper dissolves from one electrode
and is deposited on the other electrode. A non-uniform spatial distribution of
copper ion concentration is responsible for buoyancy and convection. In order
for copper ions to behave in the same way as salt or temperature in a Rayleigh-B\'enard
experiment, one needs to make sure that the ions diffuse and are advected, but
do not experience a force in the electrical field due to the potential
difference applied between the two electrodes. This is achieved by dissolving a
large concentration of another electrolyte such as $H_2SO_4$ which does not
participate in the chemical reaction at the electrodes and which screens the
electrical field in the bulk of the cell. The electrical field separates ions of
the electrolyte which accumulate in charged layers of microscopic
thickness next to each electrode so that the copper ions move essentially in a field free
environment. Refs. \onlinecite{Goldst90,Goldst92} provide more details about the
electrochemical system and its relation to Rayleigh-B\'enard convection.

In the experiments presented here, the two copper electrodes are regulated in
temperature, with a cold cathode at the bottom and a warm anode at the top.
Temperature is thus stabilizing and the ion concentration is destabilizing. Mass
transport is conveniently measured in this system because it is directly related
to the current flowing through the electrodes. Since the working fluid is
transparent, we also use particle image velocimetry (PIV) to measure velocity
and finger sizes. We are able to vary the chemical and thermal Rayleigh
numbers over several orders of magnitude. It will be shown below that the
thermal stratification is important for the mass transport and leads to
fingers even if the thermal Rayleigh number is much smaller than the chemical
one. The main results will be finger widths, flow velocities and mass transport
in convection with double diffusive fingers in a statistically steady state.

The next section presents the experimental apparatus and methods. The results
are summarized in table \ref{table1} and discussed in the third section.

\section{Experimental procedures}

This section will describe a novel experimental realization of a double
diffusive convection cell in which temperature and ion concentration are the two
diffusers. It is useful to first introduce some nomenclature and definitions to
describe the system. Double diffusive systems are characterized by four control
parameters: There are two diffusivity ratios, the Prandtl number $Pr$ and the
Schmidt number $Sc$,
\begin{equation}
Pr=\frac{\nu}{\kappa} ~~~,~~~ Sc=\frac{\nu}{D}
\end{equation}
in which $\nu$ stands for the kinematic viscosity of the convecting fluid,
$\kappa$ for its thermal diffusivity, and $D$ for the ion diffusion constant.
There are furthermore the thermal and chemical Rayleigh numbers, $Ra_T$ and
$Ra_c$,
\begin{equation}
Ra_T=\frac{g \alpha \Delta T L^3}{\kappa \nu}  ~~~,~~~ 
Ra_c=\frac{g \beta \Delta c L^3}{D \nu}
\end{equation}
with $g$ the gravitational acceleration and $L$ the cell height. The two
expansion coefficients $\alpha$ and $\beta$ determine variations of density
$\rho$ around a reference state with density, temperature, concentration and
pressure $\rho_0$, $T_0$, $c_0$ and $p_0$ via
\begin{equation}
\alpha=-\frac{1}{\rho_0}\left(\frac{\partial \rho}{\partial T} \right)_{c_0,\rho_0,p_0}
~~~,~~~
\beta=\frac{1}{\rho_0}\left(\frac{\partial \rho}{\partial c} \right)_{T_0,\rho_0,p_0}.
\end{equation}
Both $\alpha$ and $\beta$ are positive. The sign convention for the Rayleigh
numbers used in this paper is such that negative Rayleigh numbers indicate
a stable stratification. This implies that
\begin{equation}
\Delta T = T_{\rm{bottom}}-T_{\rm{top}} ~~~,~~~ \Delta c = c_{\rm{top}}-c_{\rm
{bottom}}
\end{equation}
with the subscripts indicating the boundary at which temperature $T$ or
concentration $c$ are evaluated. Another dimensionless number in common use
which will become useful below is the density ratio $\Lambda$:
\begin{equation}
\Lambda=\frac{Ra_T}{Ra_c} \frac{\kappa}{D} = \frac{\alpha \Delta T}{\beta \Delta
c}.
\end{equation}
$\Lambda$ quantifies the ratio of thermal and chemical buoyancy forces.

The electrochemical aspects of the experiment are essentially the same as in
ref. \onlinecite{Goldst90}. The cell was made of two copper electrodes of $1cm$
thickness painted with varnish on all surfaces which were not in contact with
the electrolyte. Water from thermostats was circulated through pipes welded onto
the copper plates in order to regulate their temperature. The
sidewalls were made of four plexiglass plates $1cm$ thick epoxied together to
form a rectangular frame. Four of these frames have been used, all with a cross
section of $20 cm \times 20 cm$ and heights of $L=2cm$, $4cm$, $8cm$ and $20cm$.

The cell was filled with a solution of $CuSO_4$ in 1 molar sulfuric acid. The
solution was prepared by dissolving $CuSO_4 \cdot (H_2O)_5$ at a few tens of $mmol/l$.
In between measurements, samples of the solution were extracted from the cell in
order to check with optical absorption spectroscopy that the
copper concentration remained constant.
For PIV measurements, polyamide particles $100 \mu m$ in
diameter were added to the solution. The material constants necessary for the
determination of $Pr$, $Sc$, $Ra_T$ and $Ra_c$ were taken from ref. \onlinecite{Goldst92}.

The concentration difference $\Delta c$ is not directly under control in these
experiments.
It is known only for suitable settings of the electrolytic cell.
If no current flows through the cell, $\Delta c=0$ and the concentration is
everywhere equal to the average concentration $c_0$. At small voltages applied
to the electrodes, the current rises for increasing voltage. Only diffusion
transports the copper ions close to the boundaries so that a concentration gradient
must exist near each electrode. The copper ions are depleted near the cathode
and their concentration is increased near the anode. The so called limiting
current is reached when the copper ion concentration at the cathode is down to
zero \cite{Probst95}. It is concluded from symmetry that the anode concentration 
is then $2 \cdot c_0$, and therefore the concentration difference
across the cell is $\Delta c = 2 \cdot c_0$.
Increasing the voltage further cannot increase the current any more and
a plateau is reached in the characteristic curve of current versus voltage. An
increase beyond the limiting current is only possible for voltages large enough
to allow another chemical reaction, such as water dissociation.

The voltage range for which the limiting current is obtained depends on the cell
height, the concentration $c_0$, and $\Delta T$.
Preliminary measurements of the voltage characteristic for $\Delta T = -4\,K$ 
and $\Delta T = 0$ were performed for
all heights and concentrations in order to determine the
appropriate voltage used later to create and maintain the concentration
difference of $\Delta c = 2 \cdot c_0$.

The Sherwood number, which is the same thing as the chemical Nusselt number, is
directly proportional to the number of ions transported from top to bottom
divided by the purely diffusive current, so that the Sherwood number can be
determined by
\begin{equation}
    Sh=\frac{j\,L}{z\,F\,D\,\Delta c}
\end{equation}
if $j$ is the current density, $z$ the valence of the ion ($z=2$ for $Cu^{2+}$)
and $F$ Faraday's constant. A transference number, which must sometimes be taken
into account when computing Sherwood numbers \cite{Goldst90} is so small in the
present case that it was neglected.

\begin{figure}
\includegraphics[width=8cm]{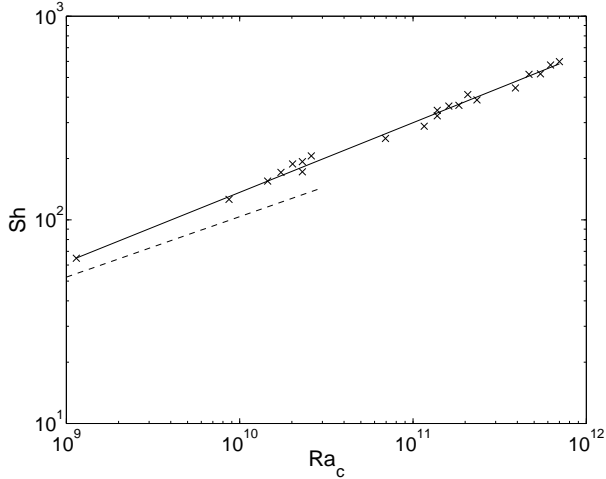}
\caption{$Sh$ as a function of $Ra_c$ in a fluid with uniform temperature. The dashed
line indicates $Sh=0.14 Sc^{-0.03}Ra_c^{0.297}$, which is the functional
dependence proposed in \onlinecite{Xia02}. The continuous line is given by
$Sh=0.052 Ra_c^{0.34}$.}
\label{fig_Xia}
\end{figure}

The double diffusion experiment was preceeded by another experiment in which the
$CuSO_4$ solution was kept isothermal in cubes of side length $5 cm$, $10
cm$ and $15 cm$. The dependence of $Sh$ on $Ra_c$ obtained from this preliminary
experiment is shown in fig. \ref{fig_Xia}. This figure contains also one point
at $Ra_c \approx 10^9$ obtained in one of the cells later used for the double
diffusive experiments (the one of height $2cm$). These results should be identical to
measurements of the Nusselt number in a thermal convection experiment. Figure
\ref{fig_Xia} also shows the $Sh(Ra_c)$ dependence extrapolated from
measurements in ref. \onlinecite{Xia02}, performed in a cylindrical cell at Rayleigh
numbers between $2 \times 10^7$ and $3 \times 10^{10}$ and Prandtl numbers
between 4.3 and 1350. In the interval of Rayleigh numbers where both experiments
overlap, there is a discrepancy of about 20\%. This may be due to the different
cell geometries, but discrepancies of this order of magnitude are not unusual
when comparing different Rayleigh-B\'enard experiments at these Rayleigh numbers
\cite{Ahlers09}.\\

\begin{figure}
\includegraphics[width=8cm]{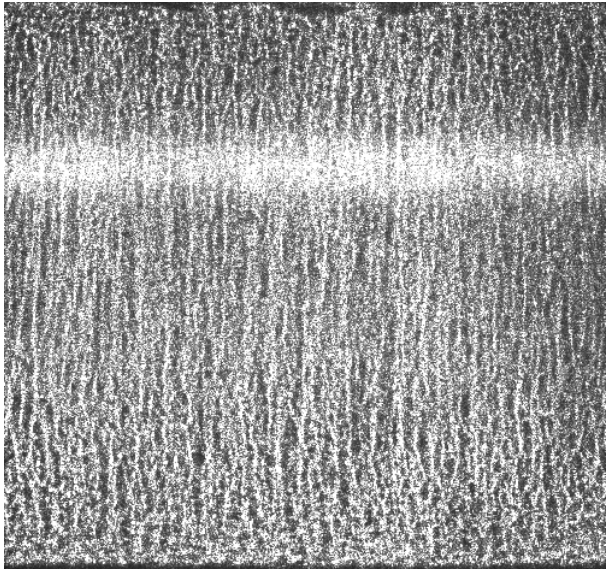}
\caption{Shadowgraph picture taken across the cell of height $80mm$ $15min$
after the voltage was switched on for $Ra_c=1.15 \times 10^{11}$ and
$Ra_T=-9.70 \times 10^7$. The bright horizontal bar is an artefact of the
illumination. The picture covers the entire height of the cell and a width of
$85mm$.}
\label{fig_shadow}
\end{figure}

\begin{figure}
\includegraphics[width=14cm]{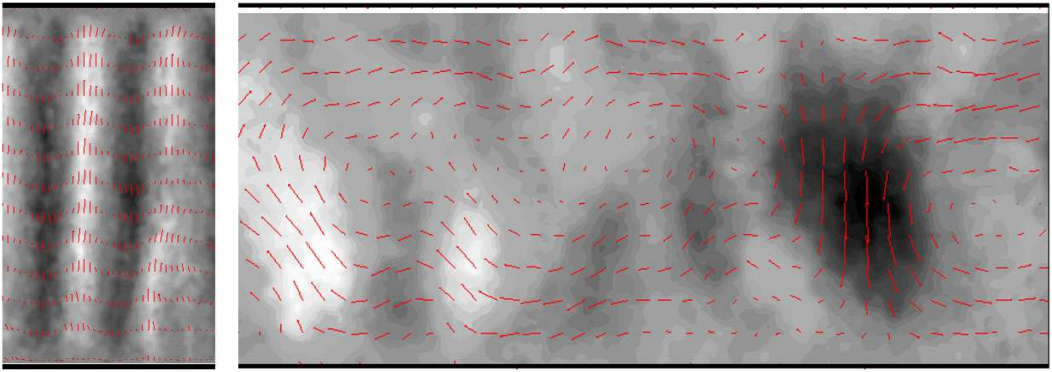}
\caption{Velocity field obtained from PIV measurements at $Ra_c=1.05 \times
10^9$ and $Ra_T=-3.81 \times 10^5$ (left panel) and $Ra_c=1.14 \times
10^9$ without any stabilizing temperature gradient (right panel).
The picture shows the entire cell height of $20mm$ and extends over a width of
$12mm$ (left) or $45mm$ (right). The lines (red online) indicate 
direction and velocity of the flow, the shade of gray depends on the vertical component
of velocity
which varies from $-0.2mm/s$ to $0.2mm/s$ (left) and from
$-0.7 mm/s$ to $0.7 mm/s$ (right) in going from dark to bright. The
left panel shows the average of 60 pictures taken during an interval of $15 s$ in
order to reduce the noise contained in any single PIV velocity field, and the right panel
shows an average over 30 pictures.}
\label{fig_PIV}
\end{figure}

A typical experimental run of double diffusion convection
was performed as follows: Care was taken when filling
the container that no air was trapped inside the cell. Then the top and
bottom plates were set to the desired temperatures and enough time was allowed
for a linear temperature gradient to establish. The voltage was then applied
instantaneously and the current observed. During transients lasting from a
few minutes to one hour, the current
oscillated with decreasing amplitude and period until the limiting current was
reached. After that time, the current fluctuated by less than 1\% and
a statistically stationary finger pattern filled the whole
cell. This pattern could be observed in shadowgraph pictures taken in parallel,
an example of which is shown in figure \ref{fig_shadow}.
The precise nature of the initial
conditions turned out to be irrelevant for the final state. A few runs were
started from uniform temperature and copper concentration, and the water
circulation through the thermostats and the voltage were switched on
simultaneously. This procedure led to longer transients but did not change the
measurements thereafter.

PIV measurements were performed in the stationary state. The
light beam of a $50mJ$ Nd:YAG laser system with two IR laser heads illuminated
a vertical plane with a width of approximately $3mm$ in the middle of the cell.
The time separation between the two pulses varied between $50$ and $1000ms$
depending on the Rayleigh numbers.
A camera took pictures of an approximately square region extending from the top
to the bottom electrode.
An example of a velocity field deduced from PIV measurements is shown in figure
\ref{fig_PIV}. Once the fingers were established, they remained surprisingly
steady, keeping their positions for $30 min$ or longer.
For each velocity field the $rms$ velocity was
calculated and averaged over all pictures to give the velocity $V$. From there, 
the non-dimensional measure of the velocity, the Reynolds number $Re$, is
computed as
\begin{equation}
Re=\frac{VL}{\nu}.
\end{equation}
For the determination of the finger size, the vertical velocities in the PIV
velocity fields were Fourier transformed along the horizontal and the spectra
averaged over all horizontal lines and all pictures. The location of the
maximum of the spectrum yields the finger thickness.
Independently, the finger thickness could simply be measured on the pictures
with a ruler.

The accuracy of the determination of $d$ is limited by the number of fingers
visible on any one picture. The relative error on $d$
is typically 10\%. The error on $V$ depends on the setting of the PIV
system and could be as large as 20 \%. The limiting current is the most accurate
of our measurements with an error of a few percent. There is also an uncertainty
on the control parameters. The measurements at small $\Delta T$ suffer most from
fluctuations in the temperature regulation. Moreover, the material constants are based on
extrapolations from relatively few measurements \cite{Goldst92} which adds up to
an uncertainty on the Rayleigh numbers on the order of 10 \%.

\newpage
\clearpage

\begin{table}\centering
\begin{tabular}{|c|c|c|c|c|c|c|c|c|c|}
\hline
L\,$[cm]$ & $\frac{\Delta T}{L}\,[\mathrm{K/cm}]$ &
$c_0\,[\mathrm{mol/l}]$ & $Pr$ & $Sc$ & $Ra_T$ & $Ra_c$ & $\frac{d}{L}$
& $Re$ & $Sh$ \\ \hline
\hline
 \vspace*{-3.3mm}
 2.0 & -10.00 &   0.025 &  8.8 & 2047.9 & $-4.18 \cdot 10^{6}$ & $+9.15
\cdot 10^{8}$ &  0.050 & $ 0.452$ & $  34.9$\\
 \vspace*{-3.3mm}
 2.0 & -10.00 &   0.030 &  8.9 & 2057.3 & $-4.18 \cdot 10^{6}$ & $+1.10
\cdot 10^{9}$ &  0.065 & $ 0.819$ & $  56.7$\\
 \vspace*{-3.3mm}
 2.0 & -10.00 &   0.034 &  9.3 & 2276.4 & $-3.80 \cdot 10^{6}$ & $+1.25
\cdot 10^{9}$ &  0.060 & $ 0.799$ & $  49.0$\\
 \vspace*{-3.3mm}
 2.0 & -10.00 &   0.044 &  8.9 & 2083.7 & $-4.16 \cdot 10^{6}$ & $+1.61
\cdot 10^{9}$ &  0.055 & $ 0.756$ & $  58.0$\\
 \vspace*{-3.3mm}
 2.0 & -10.00 &   0.055 &  9.0 & 2104.8 & $-4.15 \cdot 10^{6}$ & $+2.01
\cdot 10^{9}$ &  0.060 & $ 0.709$ & $  45.5$\\
 \vspace*{-3.3mm}
 2.0 & -5.00 &   0.015 &  8.8 & 2029.4 & $-2.10 \cdot 10^{6}$ & $+5.49
\cdot 10^{8}$ &  0.075 & $ 0.463$ & $  32.2$\\
 \vspace*{-3.3mm}
 2.0 & -5.00 &   0.023 &  8.8 & 2044.2 & $-2.09 \cdot 10^{6}$ & $+8.41
\cdot 10^{8}$ &  0.070 & $ 0.608$ & $  51.6$\\
 \vspace*{-3.3mm}
 2.0 & -5.00 &   0.031 &  8.9 & 2059.1 & $-2.09 \cdot 10^{6}$ & $+1.13
\cdot 10^{9}$ &  0.085 & $ 1.040$ & $  61.2$\\
 \vspace*{-3.3mm}
 2.0 & -5.00 &   0.037 &  9.3 & 2283.1 & $-1.90 \cdot 10^{6}$ & $+1.37
\cdot 10^{9}$ &  0.065 & $ 1.710$ & $  50.2$\\
 \vspace*{-3.3mm}
 2.0 & -5.00 &   0.044 &  8.9 & 2083.7 & $-2.08 \cdot 10^{6}$ & $+1.61
\cdot 10^{9}$ &  0.080 & $ 1.120$ & $  62.2$\\
 \vspace*{-3.3mm}
 2.0 & -5.00 &   0.052 &  9.0 & 2100.0 & $-2.08 \cdot 10^{6}$ & $+1.92
\cdot 10^{9}$ &  0.070 & $ 1.450$ & $  56.0$\\
 \vspace*{-3.3mm}
 2.0 & -3.00 &   0.015 &  8.8 & 2029.4 & $-1.26 \cdot 10^{6}$ & $+5.49
\cdot 10^{8}$ &  0.095 & $ 0.789$ & $  38.7$\\
 \vspace*{-3.3mm}
 2.0 & -3.00 &   0.026 &  8.8 & 2050.7 & $-1.25 \cdot 10^{6}$ & $+9.69
\cdot 10^{8}$ &  0.085 & $ 1.100$ & $  56.8$\\
 \vspace*{-3.3mm}
 2.0 & -3.00 &   0.033 &  8.9 & 2062.9 & $-1.25 \cdot 10^{6}$ & $+1.21
\cdot 10^{9}$ &  0.095 & $ 1.220$ & $  60.4$\\
 \vspace*{-3.3mm}
 2.0 & -3.00 &   0.044 &  8.9 & 2083.7 & $-1.25 \cdot 10^{6}$ & $+1.61
\cdot 10^{9}$ &  0.090 & $ 1.370$ & $  68.3$\\
 \vspace*{-3.3mm}
 2.0 & -3.00 &   0.050 &  9.4 & 2310.1 & $-1.14 \cdot 10^{6}$ & $+1.84
\cdot 10^{9}$ &  0.095 & $ 2.570$ & $  74.8$\\
 \vspace*{-3.3mm}
 2.0 & -1.00 &   0.016 &  8.8 & 2031.3 & $-4.19 \cdot 10^{5}$ & $+5.85
\cdot 10^{8}$ &  0.135 & $ 0.971$ & $  37.8$\\
 \vspace*{-3.3mm}
 2.0 & -1.00 &   0.024 &  8.8 & 2046.1 & $-4.18 \cdot 10^{5}$ & $+8.78
\cdot 10^{8}$ &  0.120 & $ 1.350$ & $  60.6$\\
 \vspace*{-3.3mm}
 2.0 & -1.00 &   0.029 &  9.3 & 2264.8 & $-3.81 \cdot 10^{5}$ & $+1.05
\cdot 10^{9}$ &  0.120 & $ 1.540$ & $  63.8$\\
 \vspace*{-3.3mm}
 2.0 & -1.00 &   0.032 &  9.3 & 2272.3 & $-3.80 \cdot 10^{5}$ & $+1.18
\cdot 10^{9}$ &  0.130 & $ 1.710$ & $  66.5$\\
 \vspace*{-3.3mm}
 2.0 & -1.00 &   0.033 &  8.9 & 2062.9 & $-4.17 \cdot 10^{5}$ & $+1.21
\cdot 10^{9}$ &  0.105 & $ 1.270$ & $  61.9$\\
 \vspace*{-3.3mm}
 2.0 & -1.00 &   0.042 &  8.9 & 2079.9 & $-4.17 \cdot 10^{5}$ & $+1.54
\cdot 10^{9}$ &  0.135 & $ 1.890$ & $  73.2$\\
 \vspace*{-3.3mm}
 2.0 & -1.00 &   0.048 &  9.4 & 2305.8 & $-3.79 \cdot 10^{5}$ & $+1.77
\cdot 10^{9}$ &  0.125 & $ 2.890$ & $  81.4$\\
 \vspace*{-3.3mm}
 2.0 & -0.50 &   0.030 &  9.1 & 2160.1 & $-1.99 \cdot 10^{5}$ & $+1.10
\cdot 10^{9}$ &  0.170 & $ 1.690$ & $  68.0$\\
 \vspace*{-3.3mm}
 2.0 & -0.25 &   0.031 &  8.9 & 2059.1 & $-1.04 \cdot 10^{5}$ & $+1.13
\cdot 10^{9}$ &  0.205 & $ 2.550$ & $  69.8$\\
 \vspace*{-1.5mm}
 2.0 & -0.10 &   0.030 &  9.3 & 2268.1 & $-3.81 \cdot 10^{4}$ & $+1.11
\cdot 10^{9}$ &  0.280 & $ 3.050$ & $66.2$\\
\hline
 \vspace*{-3.3mm}
 4.0 & -3.00 &   0.013 &  9.2 & 2232.3 & $-1.83 \cdot 10^{7}$ & $+3.69
\cdot 10^{9}$ &  0.045 & $ 0.652$ & $  73.3$\\
 \vspace*{-3.3mm}
 4.0 & -3.00 &   0.049 &  9.4 & 2306.7 & $-1.82 \cdot 10^{7}$ & $+1.43
\cdot 10^{10}$ &  0.045 & $ 4.410$ & $ 138.0$\\
 \vspace*{-3.3mm}
 4.0 & -2.00 &   0.014 &  9.2 & 2234.9 & $-1.22 \cdot 10^{7}$ & $+4.07
\cdot 10^{9}$ &  0.050 & $ 0.941$ & $  70.1$\\
 \vspace*{-3.3mm}
 4.0 & -2.00 &   0.050 &  9.4 & 2309.6 & $-1.21 \cdot 10^{7}$ & $+1.47
\cdot 10^{10}$ &  0.053 & $ 6.670$ & $ 141.0$\\
 \vspace*{-3.3mm}
 4.0 & -1.00 &   0.011 &  9.2 & 2229.8 & $-6.12 \cdot 10^{6}$ & $+3.33
\cdot 10^{9}$ &  0.062 & $ 1.590$ & $  91.4$\\
 \vspace*{-3.3mm}
 4.0 & -1.00 &   0.048 & 10.4 & 2871.4 & $-4.88 \cdot 10^{6}$ & $+1.43
\cdot 10^{10}$ &  0.065 & $ 6.540$ & $ 156.0$\\
 \vspace*{-3.3mm}
 4.0 & -0.50 &   0.011 &  9.2 & 2229.6 & $-3.06 \cdot 10^{6}$ & $+3.30
\cdot 10^{9}$ &  0.080 & $ 2.050$ & $  95.4$\\
 \vspace*{-1.5mm}
 4.0 & -0.50 &   0.050 &  9.4 & 2308.6 & $-3.03 \cdot 10^{6}$ & $+1.45
\cdot 10^{10}$ &  0.082 & $7.430$ & $155.0$\\
\hline
 \vspace*{-3.3mm}
 8.0 & -1.00 &   0.011 & 10.3 & 2845.6 & $-7.66 \cdot 10^{7}$ & $+2.64
\cdot 10^{10}$ &  0.030 & $ 2.180$ & $ 174.0$\\
 \vspace*{-3.3mm}
 8.0 & -1.00 &   0.049 &  9.4 & 2307.5 & $-9.70 \cdot 10^{7}$ & $+1.15
\cdot 10^{11}$ &  0.033 & $15.500$ & $ 290.0$\\
 \vspace*{-3.3mm}
 8.0 & -0.75 &   0.011 &  9.2 & 2229.2 & $-7.34 \cdot 10^{7}$ & $+2.59
\cdot 10^{10}$ &  0.035 & $ 4.380$ & $ 171.0$\\
 \vspace*{-3.3mm}
 8.0 & -0.75 &   0.050 & 10.4 & 2875.7 & $-5.85 \cdot 10^{7}$ & $+1.18
\cdot 10^{11}$ &  0.036 & $ 7.850$ & $ 291.0$\\
 \vspace*{-3.3mm}
 8.0 & -0.50 &   0.049 & 10.4 & 2874.6 & $-3.90 \cdot 10^{7}$ & $+1.17
\cdot 10^{11}$ &  0.041 & $14.600$ & $ 303.0$\\
 \vspace*{-3.3mm}
 8.0 & -0.25 &   0.011 & 10.3 & 2845.6 & $-1.92 \cdot 10^{7}$ & $+2.64
\cdot 10^{10}$ &  0.051 & $ 5.710$ & $ 185.0$\\
 \vspace*{-1.5mm}
 8.0 & -0.25 &   0.049 &  9.4 & 2307.5 & $-2.43 \cdot 10^{7}$ & $+1.15
\cdot 10^{11}$ &  0.051 & $19.100$ & $311.0$\\
\hline
20.0 & -0.50 &   0.031 &  8.9 & 2059.1 & $-2.09 \cdot 10^{9}$ & $+1.13
\cdot 10^{12}$ &  0.016 & $26.600$ & $ 608.0$
\vspace*{-3.3mm}\\
20.0 & -0.10 &   0.030 &  8.9 & 2057.3 & $-4.18 \cdot 10^{8}$ & $+1.10
\cdot 10^{12}$ &  0.028 & $55.100$ & $672.0$\\
\hline
\end{tabular}
\caption{Summary of experimental results. The first three columns give
the cell
height, the applied temperature gradient, and the concentration of
$CuSO_4$. The
next four columns contain the nondimensional control parameters (Prandtl,
Schmidt, thermal Rayleigh and chemical Rayleigh numbers) and the last
three columns the nondimensional measured quantities (finger width,
Reynolds and
Sherwood numbers).}
\label{table1}
\end{table}

\newpage
\clearpage

\section{Results and discussion}

The main quantitative results are contained in table \ref{table1}. The most
important qualitative observation from shadowgraphs and PIV is that fingers
exist at all. This is remarkable because apart from two exceptions,
$|\Lambda|<1$ in the experiments and the total density is unstably stratified so
that fingers are not necessary for convection to start. One might also think
that at $|\Lambda|=10^{-2}$, the presence of the weak stable thermal
stratification should be irrelevant for the chemical convection, but it is not.
Fingers still appear. They are of course absent for $\Lambda=0$. The isothermal experiments
in figure \ref{fig_Xia} have $\Lambda=0$ and it could be verified at the
occasion of those preliminary experiments that a single convection roll forms in a
cubic cell in this case and that the aspect ratio of convection rolls in the cell
of height $20mm$ is compatible with what is known from ordinary
Rayleigh-B\'enard convection \cite{Hartle03}.

\begin{figure}
\includegraphics[width=8cm]{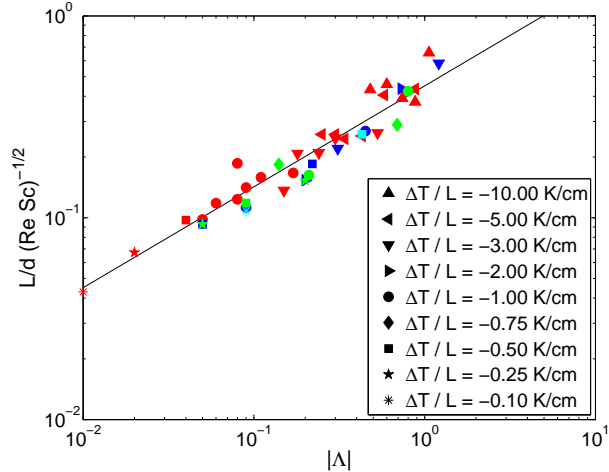}
\caption{$(Re~ Sc)^{-1/2}L/d$ as a function of $|\Lambda|$. The symbol shapes
indicate the applied temperature gradient, the colors the height of the cell:
$20 mm$ (red), $40 mm$ (dark blue), $80 mm$ (green) and $200 mm$ (light blue).}
\label{fig_diff_length}
\end{figure}

Some more information about the structure of the fingers can be deduced form
table \ref{table1}. The finger thickness always exceeds the concentration
boundary layer thickness $\lambda$, which can be computed from the measured
Sherwood number as
\begin{equation}
\lambda=\frac{L}{2 Sh}
\end{equation}
$\lambda$ is small compared with the cell height $L$ in all cases. Since the
fingers extend across the entire height of the cell, they carry fluid directly
from one boundary layer to the other. In a horizontal cross section, they must
therefore consist of a core of size $\lambda$, which is the detached boundary
layer, surrounded by the entraining fluid filling a cross section of size $d$.
The ion exchange between neighboring fingers is small as long as the distance
over which ions diffuse during the time it takes to transit from one boundary to
the other is small compared with the finger size, i.e. as long as
$(DL/V)^{1/2}/d=(Re~ Sc)^{-1/2}L/d << 1$. It can be seen from figure
\ref{fig_diff_length} that chemical exchange between fingers can be neglected for
the experiments listed in table \ref{table1}, and that this will not be true any
more around $|\Lambda| \approx 5$.

\begin{figure}
\includegraphics[width=8cm]{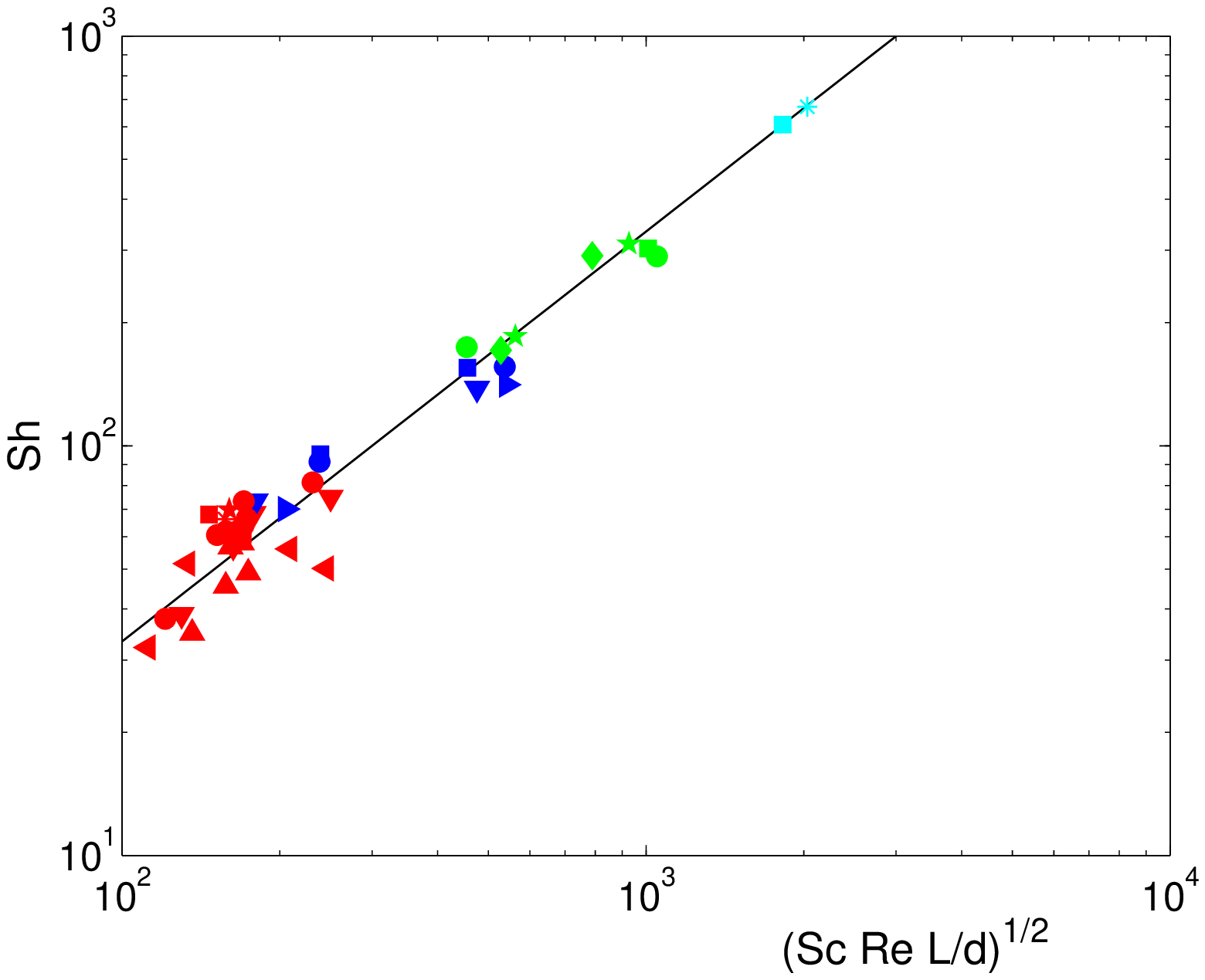}
\caption{$Sh$ as a function of $Re~ Sc~ L/d$. The line is given by eq.
(\ref{eq_Sh_Re}). The symbols are the same as in figure \ref{fig_diff_length}.}
\label{fig_Sh_Re}
\end{figure}

If there is no significant chemical diffusion across fingers, it is a simple
matter to estimate $Sh$ for $Sh >> 1$ from the advective transport:
\begin{equation}
Sh \approx Sh-1 = \frac{L}{D \Delta c} V \bar c
\end{equation}
where $\bar c$ denotes the concentration anomaly in a finger averaged over its
cross section. $\bar c=\frac{1}{2}(\lambda/d)^2 \Delta c$ in a finger with circular or square cross
section. This assumption is not compatible with the data in table \ref{table1}.
If on the other hand fingers form sheets, $\bar c =\frac{1}{2}\lambda/d \Delta c$. Figure
\ref{fig_Sh_Re} shows that
\begin{equation}
Sh = \frac{1}{3} (Re~ Sc~ L/d)^{1/2}
\label{eq_Sh_Re}
\end{equation}
is a good representation of the data from which we conclude that fingers have a
lamellar shape.

$Pr$ and $Sc$ vary by less than $30 \%$ in table \ref{table1} so that a dependence
on these parameters cannot be extracted from the data. It must be kept in mind
that the prefactors in the power laws given below potentially depend on both
$Pr$ and $Sc$.

\begin{figure}
\includegraphics[width=8cm]{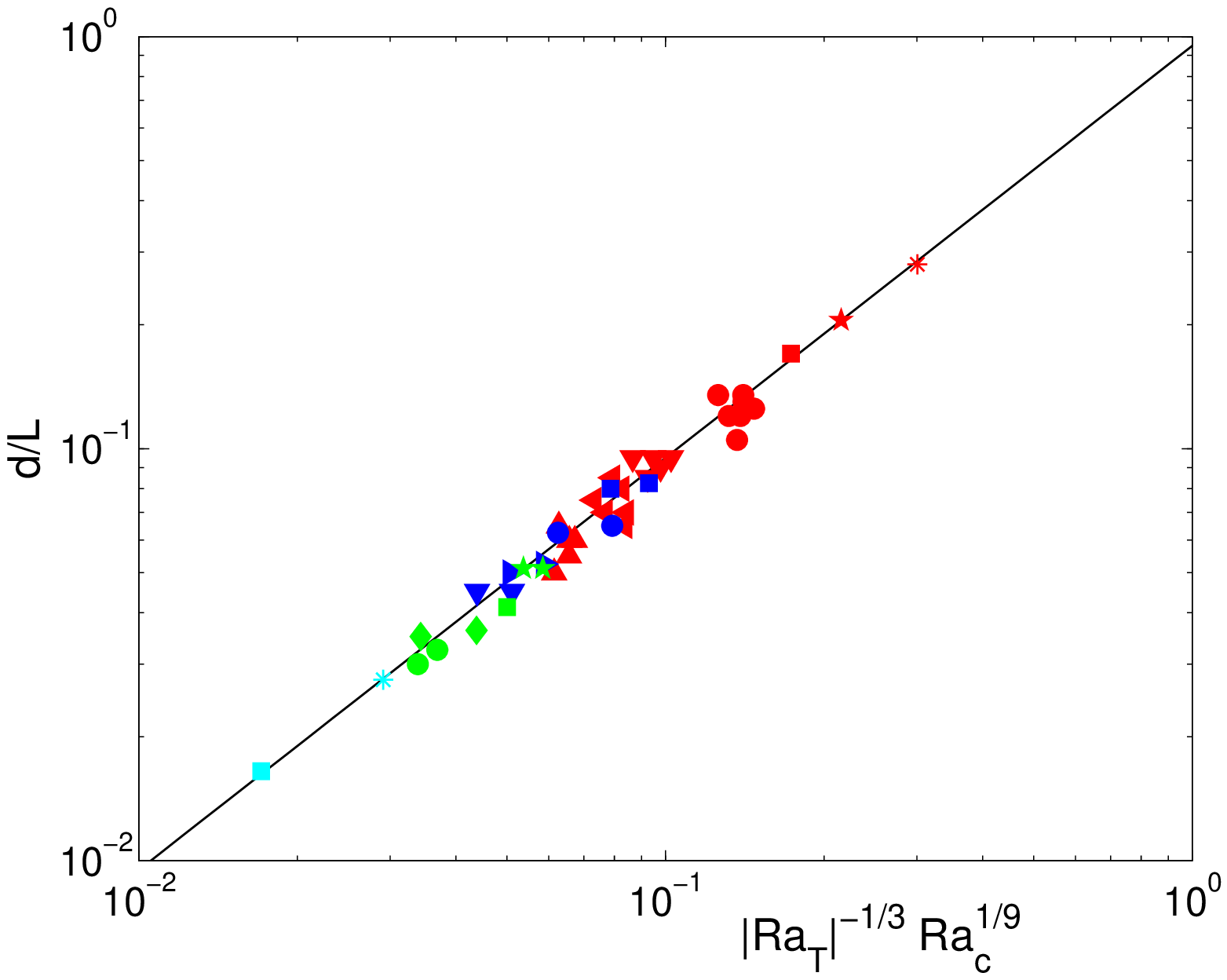}
\caption{$d/L$ as a function of $|Ra_T|^{-1/3} Ra_c^{1/9}$. The line is given by
eq. (\ref{eq_d_Ra}). The symbols are the same as in figure \ref{fig_diff_length}.}
\label{fig_d_Ra}
\end{figure}

Let us now turn to the scaling of the finger width. Which parameters do we
expect to determine this width? Long and narrow fingers should not be affected
by far away boundaries. $d$ should thus be independent of $L$. The fluid inside
fingers carries the ion concentration from the boundary layer the finger started
from. There are little diffusive losses during the transit, so that $d$ should depend
on $\Delta c$. Temperature on the other hand diffuses much more rapidly and
fingers somewhere in the bulk do not know about the boundary temperatures.
Temperature must enter the expression for $d$ through the vertical gradient,
$\Delta T/L$. A thickness $d$ which depends on $\Delta c$ and $\Delta T/L$ but
is independent of $L$ translates in nondimensional terms into a law of the form
$d/L \propto |Ra_T|^{\gamma_1} Ra_c^{\gamma_2}$ with $4\gamma_1+3\gamma_2=-1$.
The scaling known from linear stability analysis, $d/L \propto |Ra_T|^{-1/4}$
belongs to this family of power laws. However, the measured finger sizes in
table \ref{table1} depend systematically on $Ra_c$. $\gamma_1$ and $\gamma_2$
have been determined from a linear regression applied to
$\log (d/L) = A + \gamma_1 \log |Ra_T| + \gamma_2 \log Ra_c$.
Our best fit obtained with that procedure yields $\gamma_1=-0.32$ and
$\gamma_2=0.086$, which is close to
\begin{equation}
\frac{d}{L} = 0.95 |Ra_T|^{-1/3} Ra_c^{1/9}.
\label{eq_d_Ra}
\end{equation}
The quality of this fit can be judged from figure \ref{fig_d_Ra}. The important
point is that the exponents obey $4\gamma_1+3\gamma_2=-1$ so that eq.
(\ref{eq_d_Ra}) is compatible with the general picture of fingers delineated
above.

\begin{figure}
\includegraphics[width=8cm]{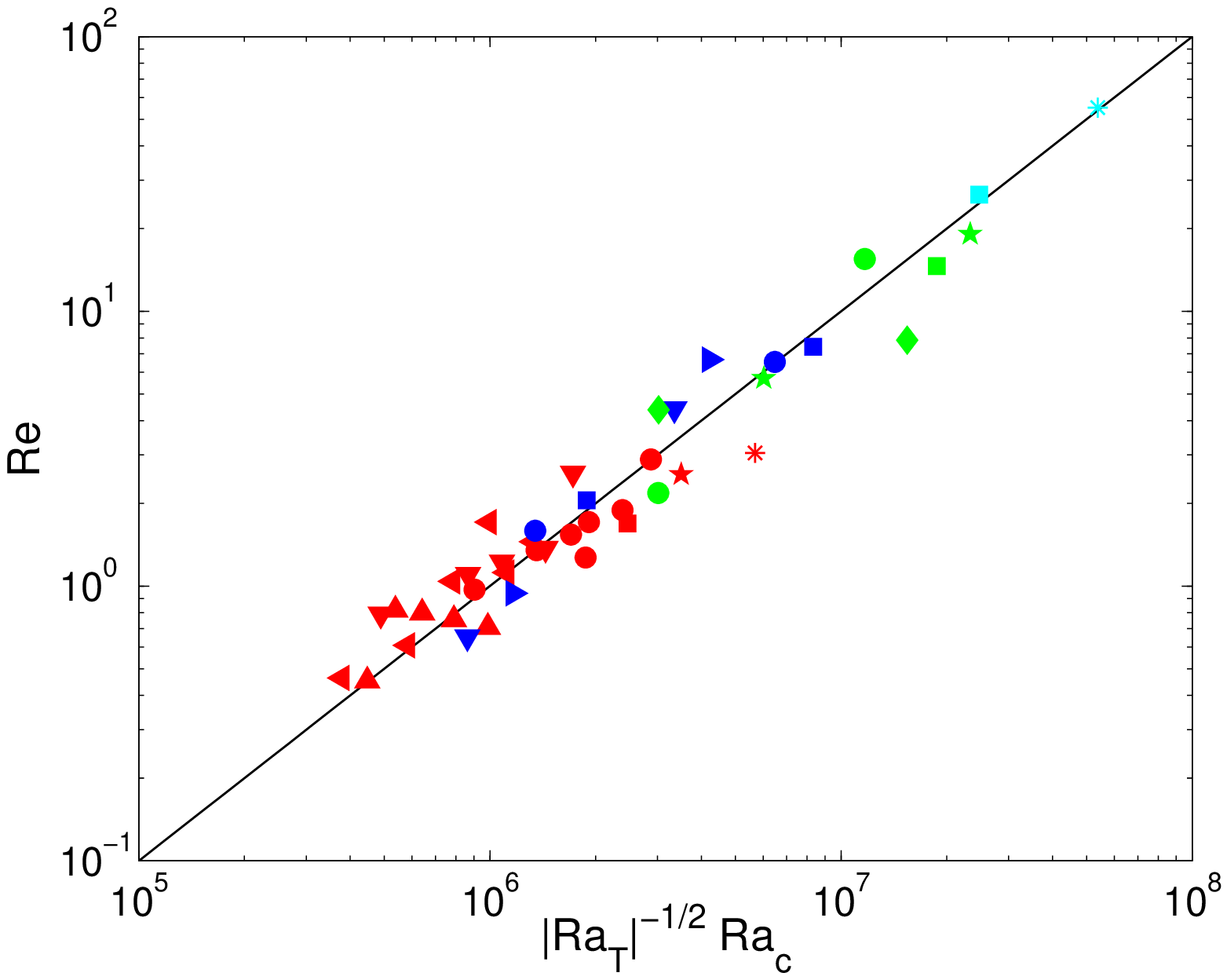}
\caption{$Re$ as a function of $|Ra_T|^{-1/2} Ra_c$. The line is given by
eq. (\ref{eq_Re_Ra}). The symbols are the same as in figure
\ref{fig_diff_length}.}
\label{fig_Re_Ra}
\end{figure}

\begin{figure}
\includegraphics[width=8cm]{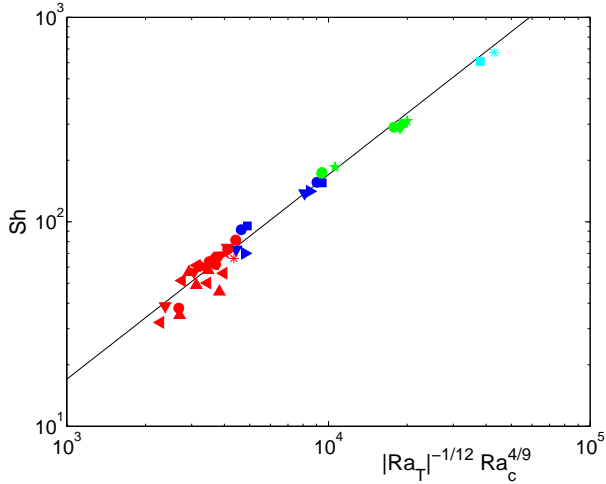}
\caption{$Sh$ as a function of $|Ra_T|^{-1/12} Ra_c^{4/9}$. The line is given by
eq. (\ref{eq_Sh_Ra}). The symbols are the same as in figure
\ref{fig_diff_length}.}
\label{fig_Sh_Ra}
\end{figure}

The velocity should in the same fashion depend on $\Delta c$, $\Delta T/L$, but
not on $L$. This means for the nondimensional variables that 
$Re \propto |Ra_T|^{\gamma_3} Ra_c^{\gamma_4}$ with $4\gamma_3+3\gamma_4=1$. An
additional argument can be made concerning the dependence on the gravitational
acceleration $g$. If buoyancy is balanced by viscous friction, one has to find
$v \propto g$. However, a balance of linear terms is questionable because the
linear terms all occur in linear stability analyses predicting fingers growing
without bounds as a function of time. Saturation of finger velocity at finite
amplitudes must involve the nonlinear terms, even if they are small. A balance
between buoyancy and the advection term leads to $v \propto g^{1/2}$. A well
known example of such a situation is Rayleigh-B\'enard convection with a
Rayleigh number near its critical value $Ra_{\rm{crit}}$. The velocity near
onset is small so that the Reynolds number is small, but experiments and weakly
nonlinear analysis \cite{Malkus58} show that velocity is proportional to 
$(Ra-Ra_{\rm{crit}})^{1/2} \propto g^{1/2}$. Returning to double diffusion, the
additional requirement $v \propto g^{1/2}$ fully determines the exponents $\gamma_3$ and
$\gamma_4$ to be $\gamma_3=-1/2$ and $\gamma_4=1$. And indeed, a good
fit to the data of table \ref{table1} is close to (see figure \ref{fig_Re_Ra})
\begin{equation}
Re=10^{-6} |Ra_T|^{-1/2} Ra_c
\label{eq_Re_Ra}
\end{equation}
A linear regression applied to the logarithm of eq. (\ref{eq_Re_Ra}) yields as
best fit
$Re=2.9 \times 10^{-6} |Ra_T|^{-0.38} Ra_c^{0.87}$. However, $Re$ is our
measurement with the largest scatter and eq. (\ref{eq_Re_Ra}) is only a
marginally worse fit than the result of the linear regression. Another reason to
prefer eq. (\ref{eq_Re_Ra}) over the direct fit is that
eqs. (\ref{eq_Sh_Re}, \ref{eq_d_Ra}, \ref{eq_Re_Ra}) now fix the dependence of
$Sh$ on $Ra_T$ and $Ra_c$ to be
\begin{equation}
Sh=0.016 |Ra_T|^{-1/12} Ra_c^{4/9}
\label{eq_Sh_Ra}
\end{equation}
which is shown in figure \ref{fig_Sh_Ra} and which is also recovered from a
direct fit by linear regression to the logarithms, whose result is
$Sh=0.017 |Ra_T|^{-0.095} Ra_c^{0.45}$.
Since $Sh \propto L/\lambda$ and since we may again argue that $\lambda$ ought
to depend on $\Delta c$, $\Delta T/L$ but not on $L$, we expect a scaling 
$Sh \propto |Ra_T|^{\gamma_5} Ra_c^{\gamma_6}$ with $4\gamma_5+3\gamma_6=1$. The
exponents in eq. (\ref{eq_Sh_Ra}) obey this constraint.
  
According to eq. (\ref{eq_Sh_Ra}), the Sherwood number for finite $|Ra_T|$
should become larger than the Sherwood number for zero temperature
stratification shown in figure \ref{fig_Xia} if $Ra_c$ exceeds $|Ra_T|$ by five
orders of magnitude. Such a combination of parameters has not been reached in
the experiments. It is not a priori impossible that the Sherwood number for
finger convection becomes larger than the Sherwood number for convection without
fingers, but it seems more likely that the corresponding combinations of control
parameters mark the limit of validity of eq. (\ref{eq_Sh_Ra}). Such a limit must
exist somewhere because $Sh \rightarrow \infty$ for $|Ra_T| \rightarrow 0$ in eq.
(\ref{eq_Sh_Ra}).

We finally compare the results of this section with previous work.
The scalings of $Sh$, $Re$ and $d/L$ show that fingers in our experiments
take the form of sheets.
On the theoretical side, as analytic treatments of the planform
selection problem became increasingly realistic, the predictions went from
square cells to rolls and back to
square cells \cite{Proctor86, Radko00}. Experimentally, both square cells
\cite{Shirtc70} and cells in the form of rolls have already been observed \cite{Hosoya89}.

Some experiments have determined a variation of finger widths with
$|Ra_T|$ compatible with $|Ra_T|^{-1/4}~$ \cite{Shirtc70, Linden73}. This is the
size of the fastest growing instability in a fluid layer with uniform
temperature and salt gradients. Starting from a step change in temperature and
salinity, ref. \onlinecite{Sreeni09} found numerically a finger width scaling as
$|Ra_T|^{-1/3}$, the same exponent that appears in eq. (\ref{eq_d_Ra})
and which is also compatible with ref. \onlinecite{Shirtc70}.
The scaling for the flow velocity, eq. (\ref{eq_Re_Ra})
has apparently never been observed before.

Past measurements of mass transport in finger convection \cite{Turner67,
Schmit79, McDoug84, Taylor89} have found good
agreement between experimental data and a relation of the form
$Sh \propto Ra_c^{1/3} f(Pr,Sc,\Lambda)$ with an undetermined function $f$,
which is sometimes called the 4/3-law because this relation states that the mass
flux varies as $\Delta c ^{4/3}$ at constant $Pr$, $Sc$ and $\Lambda$. Arguments
in favor of this type of relation come from dimensional \cite{Turner67}
and asymptotic analysis \cite{Radko00}. Eq. (\ref{eq_Sh_Ra}) can be rewritten as
$Sh \propto Ra_c^{1/3} |\Lambda|^{-1/9} |Ra_T|^{1/36}$. The variation in
$|Ra_T|^{1/36}$ is much too weak to be detectable in the experiment, so that eq.
(\ref{eq_Sh_Ra}) must be considered to be indistinguishable from the 4/3-law.

\section{Conclusion}

With an electrochemical technique, double diffusive convection can be
investigated with a tabletop experiment for $Pr \approx 9$, $Sc \approx 2200$
and chemical Rayleigh numbers ranging from $5 \times 10^8$ to $10^{12}$ while
varying the density ratio from $10^{-2}$ to 1. Fingers appear in all
circumstances. The main quantities related to mass transport, the finger
thickness, the flow velocity, and the concentration boundary layer thickness,
are all independent of the cell height $L$, and depend on the concentration
difference between top and bottom $\Delta c$ as well as on the imposed
temperature gradient $\Delta T/L$, but not on the temperature difference $\Delta
T$ alone. The scaling for the Sherwood number proposed in eq. (\ref{eq_Sh_Ra})
is experimentally indistinguishable from the classical 4/3-law. 

Perhaps the most remarkable finding is that fingers occur in the first place,
considering the small stabilizing temperature gradient used here. It remains an
open question whether there is an abrupt or a continuous transition from fingers
to ordinary convection rolls which are at least as wide as they are tall. Eq.
(\ref{eq_d_Ra}) remains correct down to the weakest stable stratification
controllable in the experiment. In the case of a continuous transition, eq.
(\ref{eq_d_Ra}) predicts that the small thermal Rayleigh number of
$|Ra_T| \approx Ra_c^{1/3}$ is
necessary for $d \approx L$. This indicates that conditions suitable for finger
formation are much more widespread than linear stability analysis suggests
\cite{Baines69}.


%

\end{document}